\documentstyle[twoside,fleqn,epsfig,espcrc2]{article}
\def\setboxz@h{\setbox\z@\hbox}
\def\wdz@{\wd\z@}

\expandafter\chardef\csname pre amssym.def at\endcsname=\the\catcode`\@
\catcode`\@=11
\def\undefine#1{\let#1\undefined}
\def\newsymbol#1#2#3#4#5{\let\next@\relax
 \ifnum#2=\@ne\let\next@\msafam@\else
 \ifnum#2=\tw@\let\next@\msbfam@\fi\fi
 \mathchardef#1="#3\next@#4#5}
\def\mathhexbox@#1#2#3{\relax
 \ifmmode\mathpalette{}{\m@th\mathchar"#1#2#3}%
 \else\leavevmode\hbox{$\m@th\mathchar"#1#2#3$}\fi}
\def\hexnumber@#1{\ifcase#1 0\or 1\or 2\or 3\or 4\or 5\or 6\or 7\or 8\or
 9\or A\or B\or C\or D\or E\or F\fi}
\font\tenmsa=msam10
\font\sevenmsa=msam7
\font\fivemsa=msam5
\newfam\msafam
\textfont\msafam=\tenmsa
\scriptfont\msafam=\sevenmsa
\scriptscriptfont\msafam=\fivemsa
\edef\msafam@{\hexnumber@\msafam}
\mathchardef\dabar@"0\msafam@39
\def\dashrightarrow{\mathrel{\dabar@\dabar@\mathchar"0\msafam@4B}}
\def\dashleftarrow{\mathrel{\mathchar"0\msafam@4C\dabar@\dabar@}}

\def\ulcorner{\delimiter"4\msafam@70\msafam@70 }
\def\urcorner{\delimiter"5\msafam@71\msafam@71 }
\def\llcorner{\delimiter"4\msafam@78\msafam@78 }
\def\lrcorner{\delimiter"5\msafam@79\msafam@79 }
\def\yen{{\mathhexbox@\msafam@55 }}
\def\checkmark{{\mathhexbox@\msafam@58 }}
\def\circledR{{\mathhexbox@\msafam@72 }}
\def\maltese{{\mathhexbox@\msafam@7A }}
\font\tenmsb=msbm10
\font\sevenmsb=msbm7
\font\fivemsb=msbm5
\newfam\msbfam
\textfont\msbfam=\tenmsb
\scriptfont\msbfam=\sevenmsb
\scriptscriptfont\msbfam=\fivemsb
\edef\msbfam@{\hexnumber@\msbfam}

\def\widehat#1{\setbox\z@\hbox{$\m@th#1$}%
 \ifdim\wd\z@>\tw@ em\mathaccent"0\msbfam@5B{#1}%
 \else\mathaccent"0362{#1}\fi}
\def\widetilde#1{\setbox\z@\hbox{$\m@th#1$}%
 \ifdim\wd\z@>\tw@ em\mathaccent"0\msbfam@5D{#1}%
 \else\mathaccent"0365{#1}\fi}
\font\teneufm=eufm10
\font\seveneufm=eufm7
\font\fiveeufm=eufm5
\newfam\eufmfam
\textfont\eufmfam=\teneufm
\scriptfont\eufmfam=\seveneufm
\scriptscriptfont\eufmfam=\fiveeufm

\catcode`\@=\csname pre amssym.def at\endcsname
\expandafter\ifx\csname pre amssym.tex at\endcsname\relax
                                         \else \endinput\fi
\expandafter\chardef\csname pre amssym.tex at\endcsname=\the\catcode`\@
\catcode`\@=11
\newsymbol\gtrsim 1326
\newsymbol\lesssim 132E
\catcode`\@=\csname pre amssym.tex at\endcsname
\def\GeV{\hbox{$\;\hbox{\rm GeV}$}}
\def\TeV{\hbox{$\;\hbox{\rm TeV}$}}
\newcommand{\Rp}{\mbox{$\not \hspace{-0.15cm} R_p$}}
\newcommand{\PTmiss}{\mbox{$\not \hspace{-0.11cm} P_{\perp}$}}
\newcommand{\picob}{\mbox{\rm ~pb$^{-1}$}}

\newcommand{\AmS}{{\protect\the\textfont2
  A\kern-.1667em\lower.5ex\hbox{M}\kern-.125emS}}
 
\title{Searches for SUSY Particles at HERA}
 
\author{Y. Sirois~\address{LPNHE, Ecole Polytechnique, IN2P3-CNRS, \\
                           91128 Palaiseau,  France}
\thanks{Invited speaker on behalf of the H1 and ZEUS Collaborations} }

\begin{document}

\begin{abstract}
\noindent
Searches at the $ep$ collider HERA for supersymmetric partners of the
Standard Model fermions are presented.
Assuming R-parity conservation, selectrons and squarks of the Minimal 
Supersymmetric Standard Model are excluded for masses up to $65 \GeV$ 
in a new region of the standard parameter space.
Admitting in addition a R-parity violating Yukawa coupling $\lambda'$, 
squarks are sought through direct $e$-$q$ fusion in a yet unexplored 
mass-coupling domain, taking into account all possible squark decays. 
Squark masses up to $240 \GeV$ are excluded for 
$\lambda' \gtrsim \sqrt{4 \pi \alpha_{em}}$,
depending only weakly on the free model parameters. 

\vspace{-9.0cm}
\begin{flushright}
  H1-10/96-494 \\
  X-LPNHE 96-08 \\
\end{flushright}
\vspace{0.8cm}  
\noindent
4th International Conference on Supersymmetries in Physics,
 \\
College Park, Maryland, USA, (May-June 1996).
\vspace{5.8cm}

\end{abstract}

\maketitle
 
\section{INTRODUCTION}
\label{sec:intro}
 
Supersymmetry (SUSY) is a likely ingredient of a true fundamental theory
beyond the Standard Model (SM).
Until the SM gets falsified by experimental data, the Minimal
Supersymmetric extension of the Standard Model (MSSM) offers a
natural framework to guide the search for new scalar superpartners
of SM fermions.

The analyses described in this paper assume the minimal
field representation of the MSSM.
This contains in particular selectrons $\tilde{e}_L$ and $\tilde{e}_R$,
and squarks $\tilde{q}_L$ and $\tilde{q}_R$, as partners of left- and
right-handed electrons and quarks.
The masses of such bosonic sparticles are treated here as free
parameters.
Gauginos and higgsinos, respectively partners of the 
$Z^0$, $\gamma$, $W^{\pm}$ gauge bosons 
and of the two required Higgs doublets,
mix in the neutral and charged sectors to form 
four neutral $\chi_{1,2,3,4}^{0}$ (neutralinos) 
and two charged $\chi_{1,2}^{\pm}$ (charginos) mass eigenstates. 
The mass mixing matrices for the $\chi_{m}^{0}$ and $\chi_{n}^{\pm}$ 
depend on the soft supersymmetry breaking parameters $M_1$ and $M_2$
associated to $SU_2$ and $U_1$ gauginos,
on the ratio $\tan \beta \equiv v_2/v_1$ of the two Higgs field vacuum 
expectation values, and finally on the mixing parameter $\mu$ associated 
to Higgs superfields.
In the strict MSSM context, a discrete symmetry ensures the
multiplicative conservation of the $R$-parity quantum number defined as
$R_p\,=\,(-1)^{3B+L+2S}$, where $B$ denotes the baryon number,
$L$ the lepton number and $S$ the spin such that $R_p = 1$ for the
SM particles and $-1$ for their superpartners.
As a consequence, sparticles must be produced by pairs and the
lightest SUSY particle (LSP) is stable.
Throughout this paper, it is assumed that the LSP is the lightest
neutralino $\chi_1^0$.
The number of free MSSM parameters is further reduced by assuming
that $M_1$ and $M_2$ are related via the weak mixing angle $\theta_W$ 
as suggested by Grand Unified Theories 
(GUT's), $M_1 = 5/3 M_2 \tan^2 \theta_W$.
No other GUT relations are used.
In particular, the mass of the gluinos $\tilde{g}$, partners of SM gluons,
is left here as a free parameter.
We assume that $M_{\tilde{g}} > M_{\tilde{q}}$ such that 
$\tilde{q} \rightarrow q \tilde{g}$ decays are kinematically forbidden.

The general SUSY superpotential allows, beyond the MSSM, for gauge
invariant terms with $R_p$ violating (\Rp) Yukawa couplings
between the quarks and leptons and their squark ($\tilde{q}$) or
slepton ($\tilde{l}$) partners.
Of particular interest for HERA are the terms 
$\lambda'_{ijk} L_{i}Q_{j}\bar{D}_k$
which allow for lepton number violating processes.
Provided that baryon number is effectively conserved at low energy,
sizeable $\lambda'$ couplings are consistent with GUT's, Supergravity 
and Superstring theories~\cite{SUPERTOE}.
They moreover do not unavoidably suffer from cosmological 
constraints~\cite{BARYSYM} and are even required for baryon asymmetry 
genesis in some cosmological models with first order electroweak phase 
transition~\cite{BARLVIO}.
By convention the $ijk$ indices correspond to the generations of
the superfields $L_{i}$, $Q_{j}$ and $\bar{D}_k$
containing respectively the left-handed lepton doublet, quark doublet
and the right-handed quark singlet.
Expanded in terms of matter fields, the corresponding interaction
Lagrangian is~\cite{BUTTER93}~:
\begin{eqnarray}
{\cal{L}}_{L_{i}Q_{j}\bar{D_{k}}}
  & = & \lambda^{\prime}_{ijk}
  \left[
   -\tilde{e}_{L}^{i} u^j_L \bar{d}_R^k - e^i_L \tilde{u}^j_L \bar{d}^k_R
    \right.
    \nonumber \\
  &   & \mbox{}
   - (\bar{e}_L^i)^c u^j_L \tilde{d}^{k*}_R
                                      + \tilde{\nu}^i_L d^j_L \bar{d}^k_R
    \nonumber \\
  &   & \mbox{}
    \left.
   + \nu_L \tilde{d}^j_L \bar{d}^k_R
                               + (\bar{\nu}^i_L)^c d^j_L \tilde{d}^{k*}_R
  \right]
   + \mbox{h.c.}           \nonumber
    \nonumber
\end{eqnarray}
where the superscripts $^c$ denote the charge conjugate spinors
and the $^*$ the complex conjugate of scalar fields.
Hence the couplings $\lambda'_{1jk}$ allow for resonant production
of squarks through $e$-$q$ fusion.

Slepton-squark searches at HERA in the MSSM framework are described in
section~\ref{sec:mssm}. Squark searches in the \Rp\ extension of the 
MSSM are discussed in section~\ref{sec:rpsusy}.
Conclusions are presented in section~\ref{sec:concl}
 
\section{SEARCH FOR SLEPTON-SQUARK PRODUCTION IN THE MSSM}
\label{sec:mssm}
 
\begin{figure}[h]
 
\vspace{3.1cm}
 
\hspace{-4.0cm}  \epsfxsize=6.0cm \epsffile{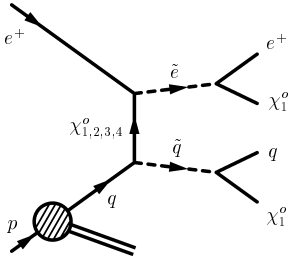}
 
\vspace{-3.5cm}

 \caption[]{ \label{fig:mssmdiag}
            {\small Lowest order Feynman diagram for $\tilde{e}+\tilde{q}$
            production in $ep$ collisions with typical subsequent sfermion 
            decays involving the LSP. }}
\end{figure}

The dominant MSSM process at HERA is the pair production
$e p \rightarrow \tilde{e} \tilde{q} X $ via $t$-channel
$\chi_m^0$ exchange as shown in Fig.~\ref{fig:mssmdiag}.
The production cross section~\cite{BARTELS88} 
depends on first approximation only on the
$\chi_1^0$ mass and couplings, and on the hard process energy threshold
$(M_{\tilde{e}}+M_{\tilde{q}})$.
The sfermion decays involve mainly the $\chi_1^0$ which lead to striking
event signatures looking like Deep Inelastic Scattering Neutral Current
(DIS NC) except for large associated missing energy and momentum.

The H1 analysis~\cite{H1MSSM} based on an integrated luminosity of 
$L = 6.4 \picob $, requires an isolated $e$ within
$10^{\circ} < \theta^e < 135^{\circ}$ possessing an energy
$E^e \geq 10 \GeV$ and a transverse momentum $P_{\perp}^e \geq 8 \GeV$.
Furthermore, it is asked that the quantity $\sum E - P_z$ which
peaks at twice the incident electron energy (i.e. $55 \GeV$) for
DIS NC events be found below $40 \GeV$.
Finally, the absolute value of both the parallel and perpendicular
(relative to the measured electron) component of the total
transverse ``momentum'' $\vec{P}_{\perp}$ should be large
($\gtrsim 3 \rightarrow 7 \GeV$).

No event candidate remains in the data while $0.6 \pm 0.2$ radiative
\begin{figure}[htb]
 
 \vspace*{-1.9cm}

 \hspace*{-0.5cm}
 \epsfig{file=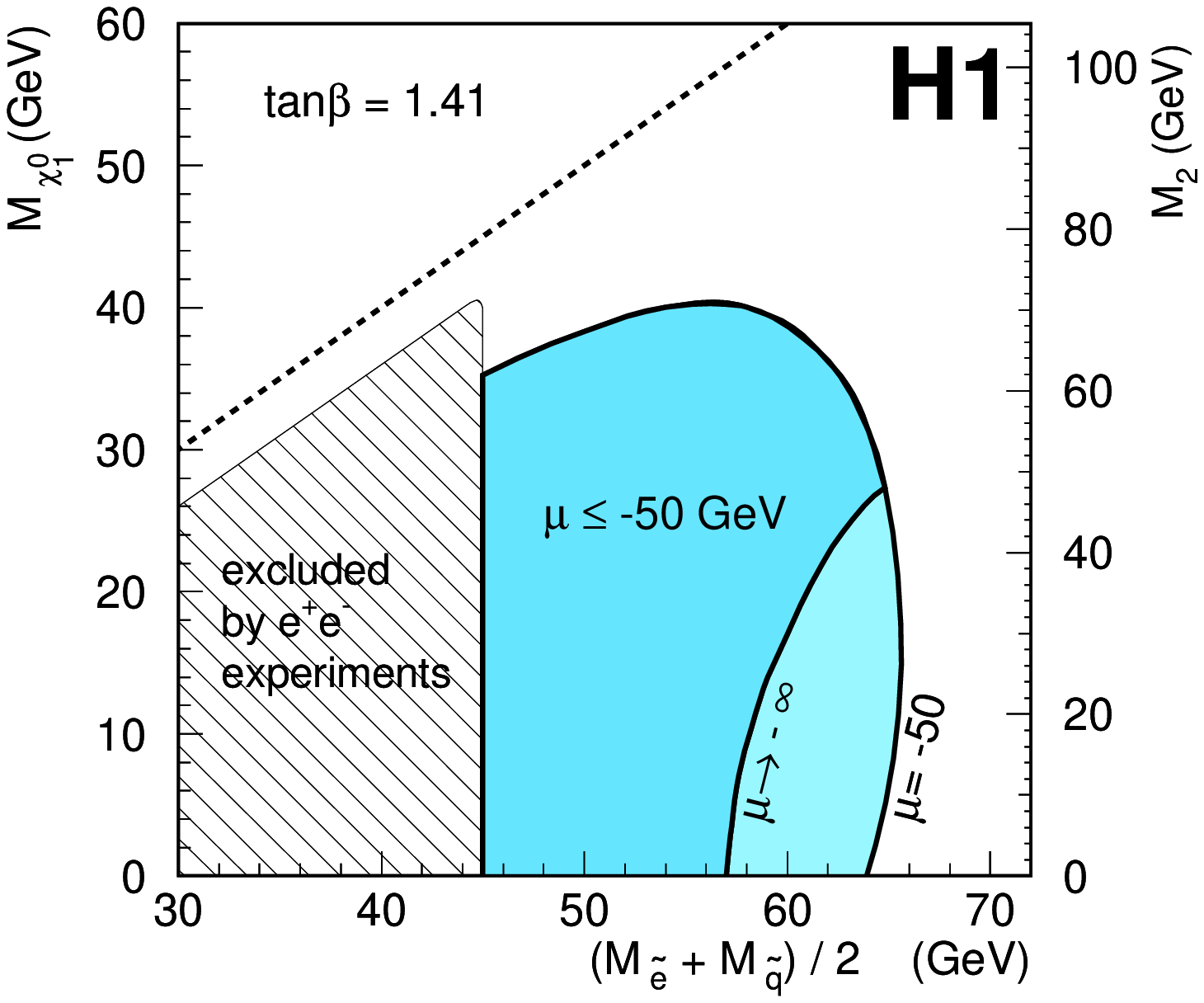,bbllx=50pt,%
                               bblly=450pt,%
                               bburx=550pt,%
                               bbury=650pt,%
                               width=8.3cm,%
                               angle=0}
 
\vspace*{2.7cm}

 \hspace*{-0.5cm}
 \epsfig{file=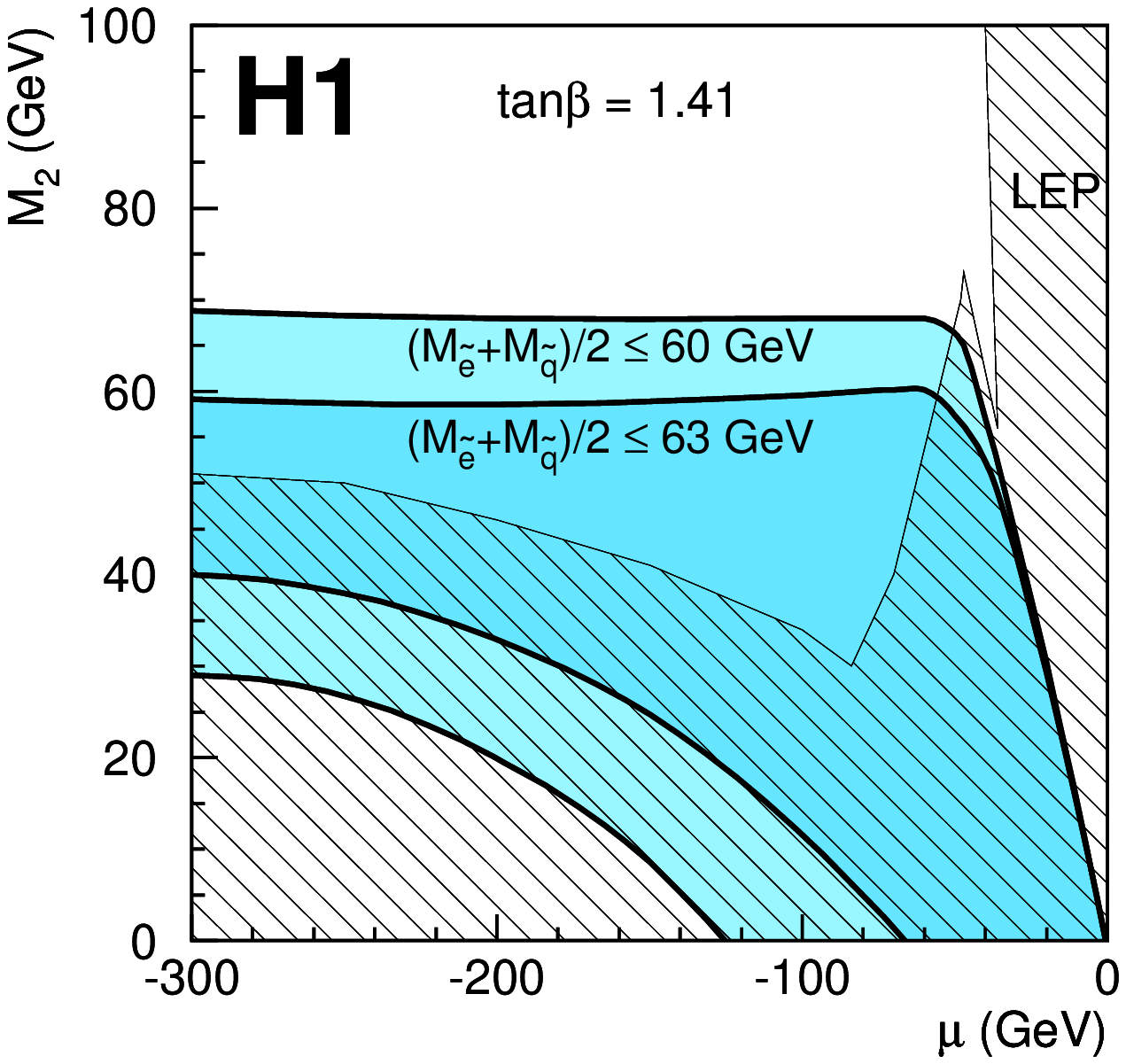,bbllx=50pt,%
                               bblly=450pt,%
                               bburx=550pt,%
                               bbury=650pt,%
                               width=8.3cm,%
                               angle=0}
\vspace*{2.3cm}
\caption[]{ \label{fig:m2mu}
          {\small Excluded mass domains at 95\% CL in the
          ({\it top}) $M_{\chi_1^0}$ versus 
          $(M_{\tilde{e}} + M_{\tilde{q}})/2$ plane and
          ({\it bottom}) in the $M_2$ versus $\mu$ plane for 
          $\tan \beta = 1.41$.}}
\end{figure}
charge current (CC) events that mimic the desired event topology are
expected. This background dominates over expectations from DIS NC
and $\gamma$-$g$ fusion (photoproduction) processes.
The signal detection efficiency $\epsilon({\cal{P}})$ with
${\cal{P}}^2 \equiv
  (M^2_{\tilde{e}}-M^2_{\chi_1^0})(M^2_{\tilde{q}}-M^2_{\chi_1^0})
                                      /4M_{\tilde{e}}M_{\tilde{q}}$
is estimated to reach $56\%$ for ${\cal{P}} > 25 \GeV$ and drops to
$\lesssim 20\%$ for ${\cal{P}} \lesssim 10 \GeV$.
 
Exclusion limits are derived at 95\% Confidence Level (CL) and presented 
in Fig.~\ref{fig:m2mu}.
The excluded mass range (Fig.~\ref{fig:m2mu} {\it top}) extends to
$(M_{\tilde{e}}+M_{\tilde{q}})/2 = 65 \GeV$ and to
$M_{\chi_1^0} = 40 \GeV$.
The search is sensitive down to mass differences of
$M_{\chi_1^0} - M_{\tilde{e},\tilde{q}} \sim 10 \GeV$.
In the MSSM parameter space, the excluded domain 
(Fig.~\ref{fig:m2mu} {\it bottom}) covers an area at $\mu \ll 0$ where the 
$\chi_1^0$ is dominated by its $\tilde{\gamma}$ component 
such that the couplings of the $\tilde{e}$ and $\tilde{q}$ are of
electromagnetic strength and the cross section is maximal.
The domain excluded by H1 results extends to larger 
values of $M_2$ than recent MSSM searches at LEP results~\cite{LEPMSSM}
and might still be uncovered by searches at the TEVATRON 
for $M_{\tilde{g}} \gg M_{\tilde{q}}$ and 
$M_{\chi_1^0} \sim M_{\tilde{q}}$~\cite{SCHLEPER}.

\section{SEARCH FOR \Rp-SUSY SQUARKS}
\label{sec:rpsusy}
 
\begin{figure}[htb]
\vspace{-1.3cm}
 
\epsfxsize=12.0cm \epsffile{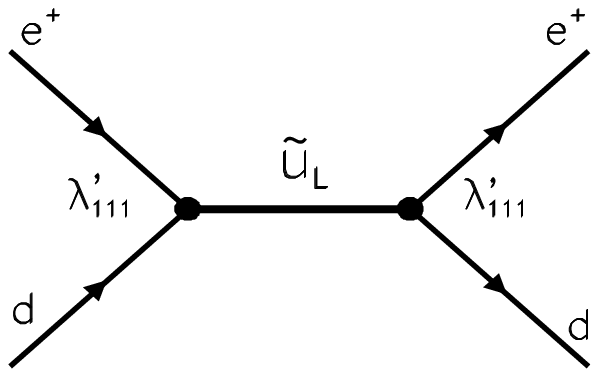}
\vspace{-8.8cm}
 
\hspace*{0.2cm}
\epsfxsize=12.0cm \epsffile{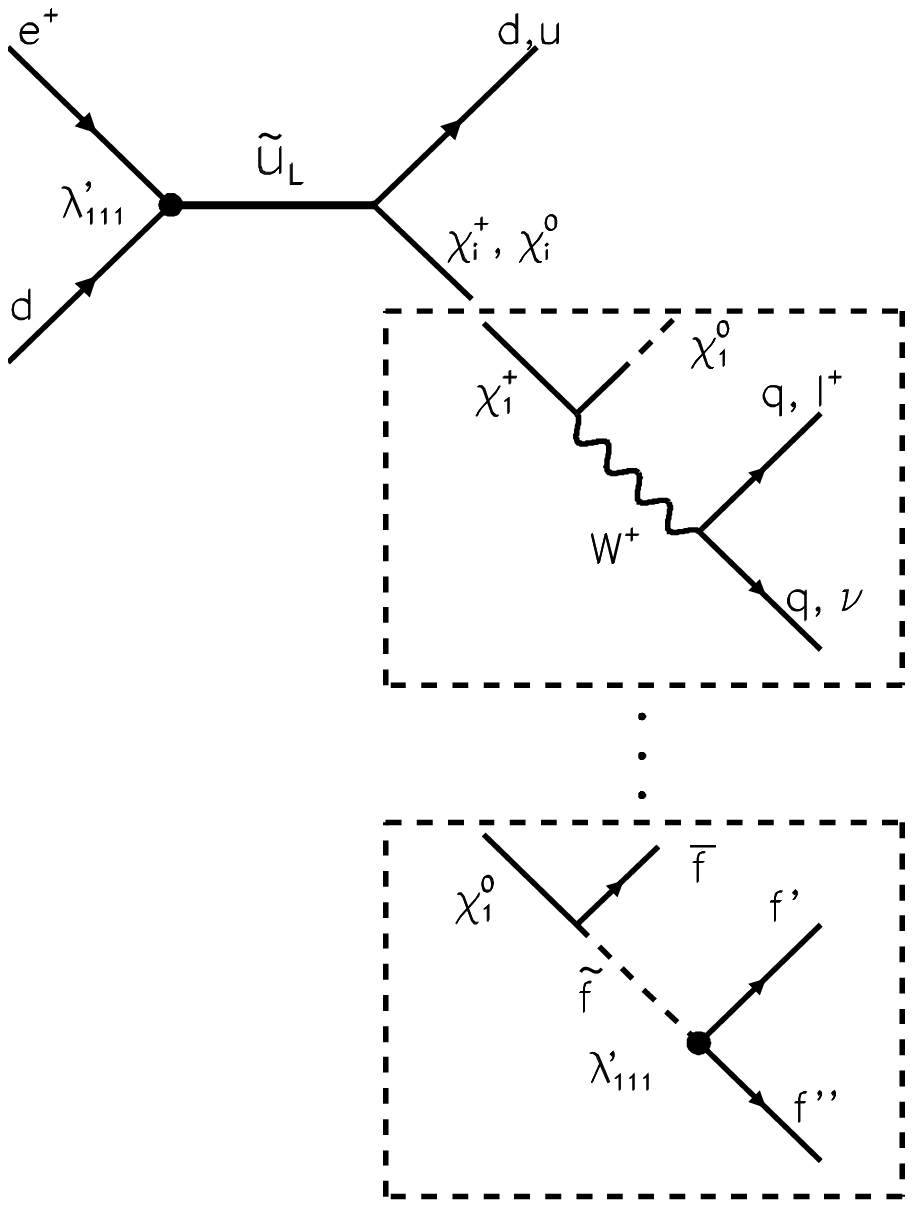}
\vspace{-4.3cm}

 \caption[]{ \label{fig:sqdiag}
      {\small Lowest order $s$-channel diagrams for $\tilde{u}_L$ squark
      production at HERA followed by 
      ({\it top}) a \Rp\ decay or 
      ({\it bottom}) a gauge decay involving a gaugino-higgsino $\chi$. 
      Typical decays of the emerging $\chi_1^0$
      or $\chi_1^+$ are shown in the dotted boxes. }}
\end{figure}

 
Squarks can be directly produced in the $s$-channel at HERA through $e$-$q$
fusion as shown in Fig.~\ref{fig:sqdiag}.
By gauge symmetry, only partners of left-handed $u$-like quarks
(i.e. $\tilde{u}_L$, $\tilde{c}_L$ or $\tilde{t}_L$) or right-handed $d$-like
quarks (i.e. $\tilde{d}_R$, $\tilde{s}_R$ or $\tilde{b}_R$) can be resonantly
produced through nine possible $\lambda'_{1jk}$ couplings.
The production cross-section scales with $\lambda'^2$ and with the probability
$q'(x)$ to find in the proton the relevant quark with momentum fraction
$x \simeq M^2_{\tilde{q}} / s_{ep}$ for $ep$ collisions at
$\sqrt{s_{ep}} \simeq 300 \GeV$.
The couplings $\lambda'_{11k}$ are best probed with an $e^-$ beam via
processes involving the $u$ valence quark
(e.g. $e^- + u \rightarrow \tilde{s}_R$) while $\lambda'_{1j1}$ ($j \neq 1$)
are best probed with an $e^+$ beam via processes involving the $d$ valence
quark  (e.g. $e^+ + d \rightarrow \tilde{c}_L$).
The other $\lambda'_{1jk}$ couplings ($j,k \neq 1$) involve heavier quark
flavours that have to be extracted from the sea.

 
The dominant squark decays result from a competition between the
\Rp\ Yukawa couplings and the MSSM gauge couplings.
``Right''- and ``left''-handed squarks have different allowed or dominant
decay modes. 
In ``\Rp-decay modes'', the $\tilde{u}_L$-like squarks decay only via
$\tilde{u}_L \stackrel{\lambda'}{\rightarrow} e^+ d$
while the $\tilde{d}_R$-like squarks have an additional decay with missing
transverse momentum due to an escaping neutrino
$\bar{\tilde{d}}_R \stackrel{\lambda'}{\rightarrow} e^+ \bar{u}$ or
$\bar{\nu} \bar{d}$.
The partial decay widths for these \Rp-decay modes scale with
$\lambda'^2 \times M_{\tilde{q}}$.
In ``gauge decay'' modes, the $\tilde{u}_L$-like decays involve either
the $\chi_m^0$ or the $\chi_n^+$,
$\tilde{u}_L \rightarrow u \chi_m^0$ or $d \chi_n^+$.
In contrast the gauge symmetry forbids an electroweak coupling of the
$\tilde{d}_R$-like squarks to the $\chi_n^+$.
Hence the $\tilde{q}_R \rightarrow q' \chi_n^+$ is strongly suppressed
for first and second generation squarks as it can only proceed through
the $\tilde{H}^+$ component of the $\chi^+$ with a coupling proportional
to the $q'$ mass.
The partial decay widths for the gauge decays scale with
$M_{\tilde{q}} \times (1 - M_{\chi_m^0}^2 / M_{\tilde{q}}^2)^2$
and, in detail, depend on MSSM parameters via the mass-mixing in the
gaugino-higgsino sector~\cite{HERAWRK}.
The gauge decay modes are expected to dominate through most of the
parameter space for the $\lambda'-M_{\tilde{q}}$ domain accessible at
HERA. The \Rp-decay modes will only dominate near the kinematical limit
(e.g. at $M_{\tilde{q}} \gtrsim 250 \GeV$) where a large $\lambda'$ is
required at production.

\begin{figure}[htb]
\vspace{-1.0cm} 

\hspace*{0.3cm}
\epsfxsize=6.1cm \epsffile{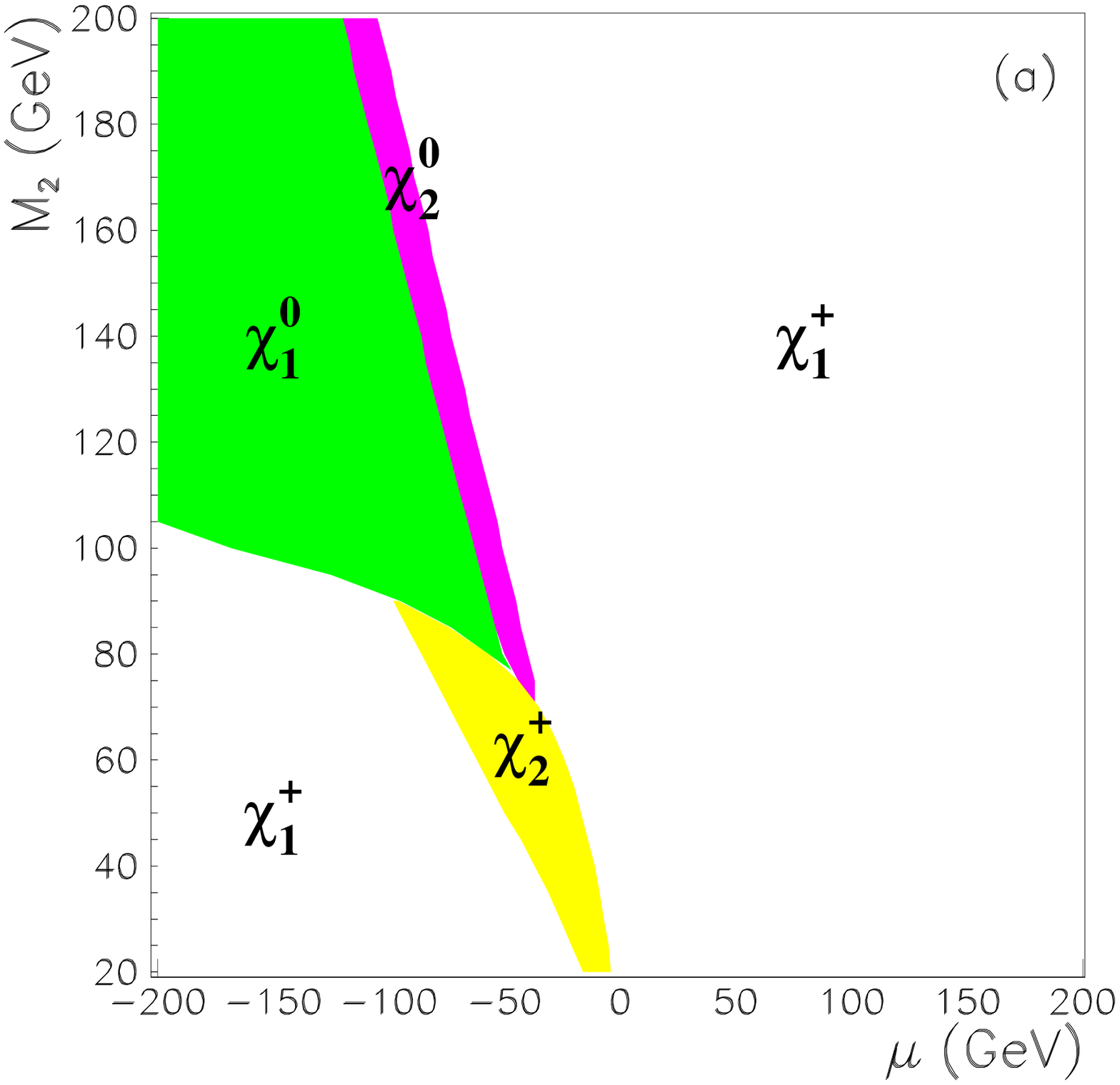}
 
\vspace{-0.2cm}
 
\hspace*{0.3cm}
\epsfxsize=6.1cm \epsffile{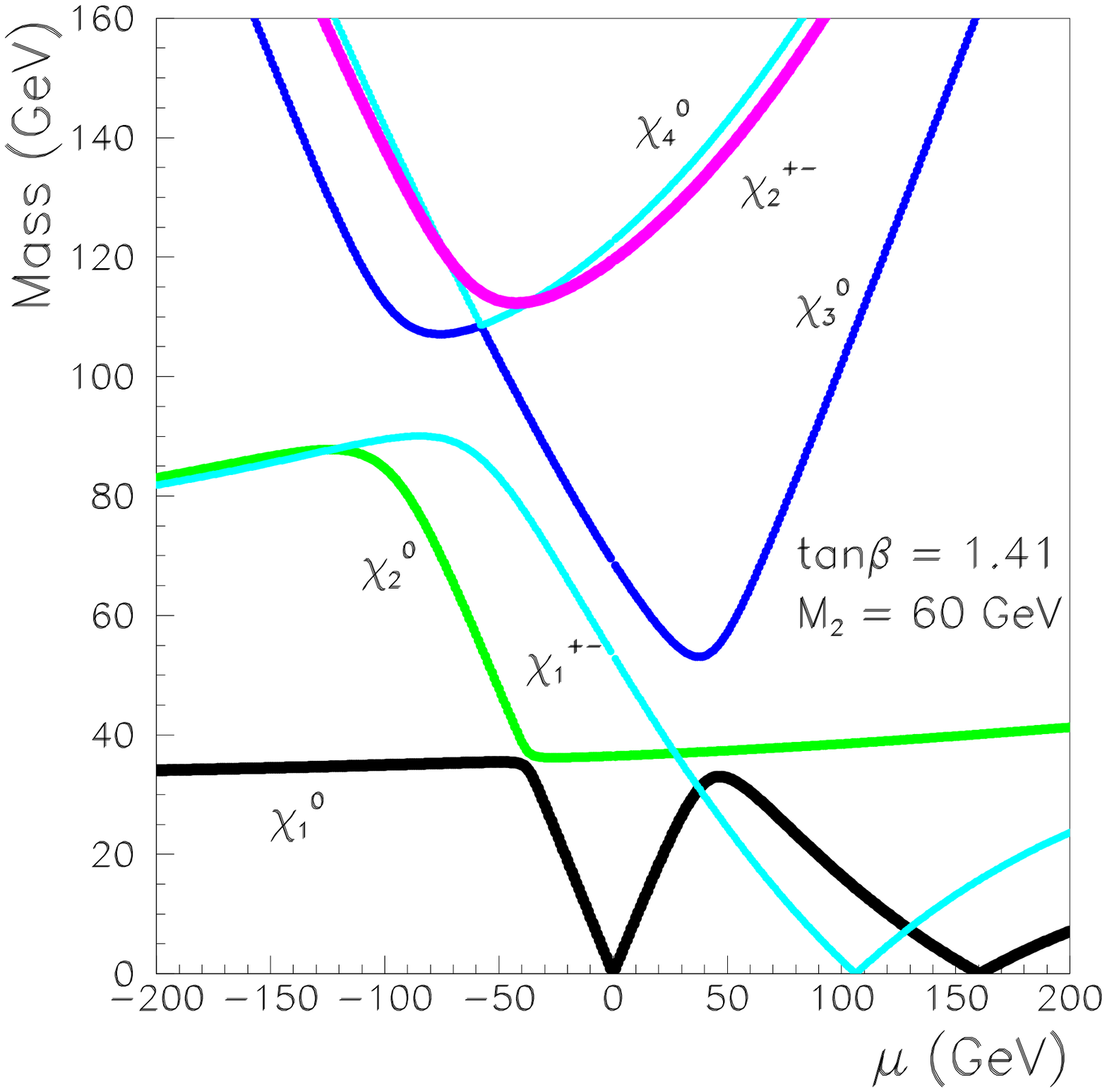}
 
\vspace{-0.8cm}

 \caption[]{ \label{fig:sulbr}
            {\small ({\it top}) $\chi_m^0$ or $\chi_n^{\pm}$
            eigenstates involved in the dominant decay modes of the
            $\tilde{u}_L$ in regions of the ($M_2, \mu$) plane for
            $\tan \beta = 1$;
            ({\it bottom}) physical masses of the $\chi_m^0$ or 
            $\chi_n^{\pm}$ versus $\mu$ for $M_2 = 60 \GeV$. }}
\end{figure}

The $\tilde{u}_L$ gauge decays dominantly involve a $\chi_n^+$ as soon 
as kinematically allowed, i.e. everywhere except in a domain at large
$M_2$ and $\mu \ll 0$ where the $\chi_m^0$ is heavy and has a large 
$\tilde{\gamma}$ component. 
This can be seen in Fig.~\ref{fig:sulbr}.
The lightest state $\chi_1^+$ is found to be concerned in a vast portion 
of the parameter space. 
The $\tilde{d}_R$ gauge decay involves dominantly the $\chi_1^0$
($> 90\%$ of gauge decays) when this LSP is dominantly
$\tilde{\gamma}$-like, i.e. at $\mu/M_2 \lesssim -5/9$.
Else the LSP participates in $20$ to $80\%$ of the gauge decays in the
region where its $\tilde{Z}$ component dominates.
Its contribution is negligible only in the $\mu \lesssim 0$ region
when dominated by its $\tilde{H}$ component.
 
The $\chi$'s couple to fermion-boson pairs.
The $\chi^+$ can decay into a $\chi^0$ and two matter fermions (quark
or lepton pairs) via the exchange of a real or virtual $W^+$ boson, e.g.
$\chi_1^+ \rightarrow \chi_1^0 W^+; W^+ \rightarrow qq'$ or
$l^+ \nu$. It can also decay into three matter fermions (a lepton and
two quarks) when involving the \Rp\ couplings $\lambda'_{1jk}$ through
the exchange of a virtual sfermion, e.g.
$\chi_1^+ \rightarrow \bar{d} \tilde{u}_L;
 \tilde{u}_L \stackrel{\lambda'}{\rightarrow} e^+ d$ or
$\chi_1^+ \rightarrow u \bar{\tilde{d}}_L;
 \bar{\tilde{d}}_L \stackrel{\lambda'}{\rightarrow} \nu \bar{d}$.
The \Rp\ decays
$\chi_1^+ \rightarrow e^+ \bar{d}_j d_k$ or $\nu_e u_j \bar{d}_k$
dominate for $\lambda'_{1jk}$ couplings values of
$\cal{O}(\sqrt{4 \pi \alpha_{em}})$.
The $\chi_1^0$ is generally unstable and will also decay via the \Rp\ coupling
$\lambda'_{1jk}$ through virtual sfermion exchange,
$\chi_1^0 \rightarrow \nu d_j \bar{d}_k$, $e^+ \bar{u}_j d_k$ or
$e^- u_j \bar{d}_k$.
The $\chi_1^0$ being a Majorana particle, there is an equal probability
for decays into the like (with respect to the incident $e$ beam)
and unlike sign leptons.
Thus, this leads to striking and background free lepton number violating
signals.
The $\chi_1^0 \rightarrow e^{\pm} q \bar{q}$ decays dominate
( 63\% to 88\% of the $\chi_1^0$ branching ratio ) where the $\chi_1^0$
is $\tilde{\gamma}$-like.
The $\chi_1^0 \rightarrow \nu q \bar{q}$ decays generally dominate where
the $\chi_1^0$ is $\tilde{Z}$-like.
A $\tilde{H}$-like $\chi_1^0$ will most probably be long-lived and
escape detection.

In summary, the event topologies can be broadly classified in three families.
(A) squark \Rp\ two-body decays (e.g. $\tilde{q}' \rightarrow \nu q'$
   or $\tilde{q}' \rightarrow e q$);
(B) squark gauge decays into three charged fermions final states
   (e.g. $ \tilde{q}' \rightarrow q' \chi_1^0 ;
                                     \chi_1^0 \rightarrow e^{\pm} q'' q'''$)
   or for $\tilde{u}_L$-like squarks,
   (e.g. $ \tilde{q}' \rightarrow q \chi_1^+ ;
                                      \chi_1^+ \rightarrow e^+ q \bar{q}$)
(C) squark gauge decays with escaping $\nu$'s (missing transverse momentum
   \PTmiss), jet(s) and/or charged lepton(s).
These were studied in a recent H1 analysis~\cite{H1RPSUSY} based on 
$L = 2.8 \picob$.
The event topologies of type (A) are indistinguishable on event-by-event
from the DIS NC or CC backgrounds. These are statistically 
suppressed by requiring either an isolated $e$ with 
$P^e_{\perp} > 7 \GeV$ in the angular range 
$10^{\circ} < \theta^e < 145^{\circ}$ ($e + q + X$ final states) or and overall 
$\PTmiss > 25 \GeV$ ($\nu + q + X$ final states). 
In the former case, the background is further
suppressed by requiring that the $e$ be found at high $y_e$ where
$y_e = 1/2 (1 + \cos \theta_e^*)$ and $\theta_e^*$ is the electron
angle in the $e$-$q$ CM frame.
The uniform decay of the scalar particle in the CM frame leads to a
flat $y_e$ spectrum which is in contrast to the $1/y_e^2$ spectrum
expected for the DIS NC background at fixed quark momentum fraction $x$.
The DIS NC background for event topologies of type (B) with a ``right''
sign $e$ is suppressed by exploiting both the angular distribution of the
$e$ and that of the highest $P_{\perp}$ jet found in the azimuthal
hemisphere opposite to the electron~\cite{WARSAW96}.
Event topologies of type (B) with a ``wrong'' sign $e$ are essentially
background free. 
\begin{figure}[htb]
\vspace{-1.1cm}
 
\hspace*{0.2cm}
\epsfxsize=7.4cm \epsffile{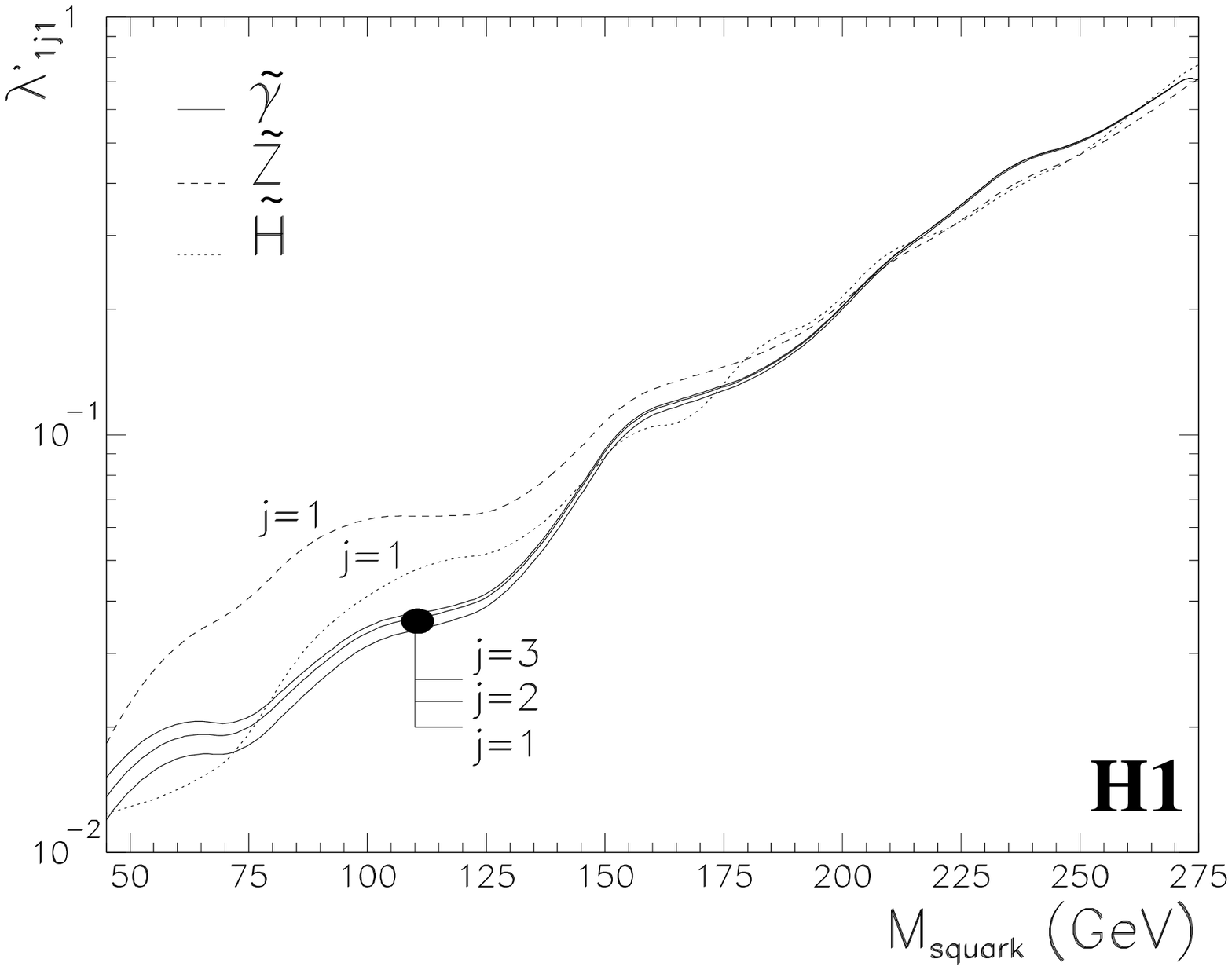}
 
\vspace{-0.5cm}
 
\hspace*{-1.7cm} \epsfxsize=10.7cm \epsffile{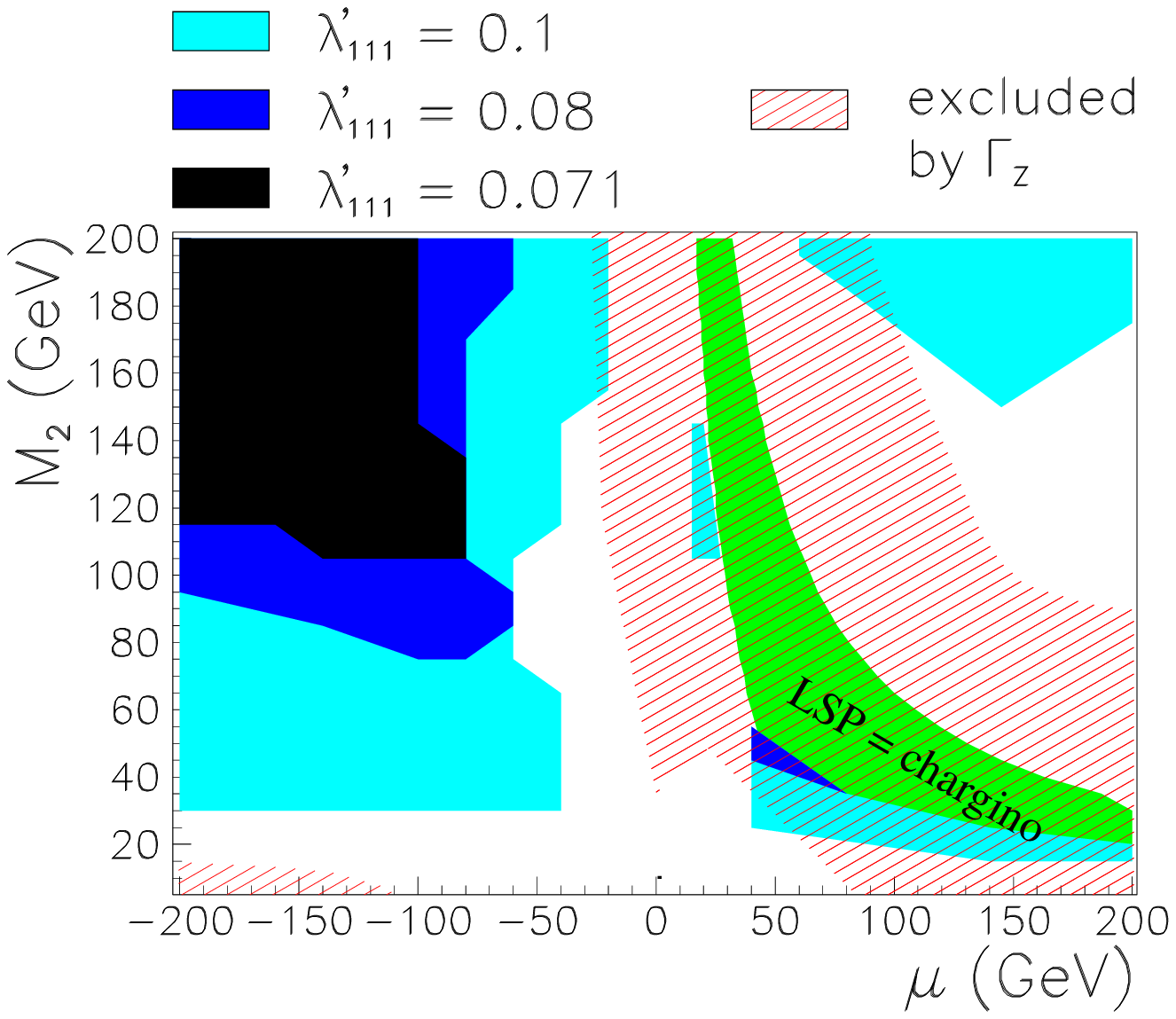}
 
\vspace{-1.4cm}

 \caption[]{ \label{fig:limplan}
           {\small ({\it top}) Exclusion upper limits at 95\% CL on 
           $\lambda'_{1j1}$ (regions above the curves excluded); 
           ({\it bottom})
           excluded domains in the $M_2$ versus $\mu$ plane
           for a $\lambda'$ close to the upper limits at 
           $M_{\tilde{q}} = 150 \GeV$. }}
\end{figure}
\begin{table}[hbt]
%
\caption{{\small Exclusion upper limits at 95\% CL on 
         $\lambda'_{1jk}$ for $M_{\tilde{q}} = 150 \GeV$ and
         $M_{\chi_1^0} = 40 \GeV$ for $\tilde{\gamma}$-like and
         $\tilde{Z}$-like $\chi_1^0$.
         Indirect limits have been scaled to $M_{\tilde{q}} = 150 \GeV$
         and are quoted at 95\% CL.}}
\label{tab:limtab}
\begin{tabular*}{7.4cm}{@{}l@{\extracolsep{\fill}}llll}
\hline
                 & \multicolumn{2}{l}{ \underline{{\bf H1} limits} }
                 & \multicolumn{2}{l}{ \underline{Indirect limits} } \\
                 & \, $\tilde{\gamma}$-like
                 & $\tilde{Z}$-like
                 & \multicolumn{1}{l}{Value}
                 & \multicolumn{1}{l}{Process \,\, [Ref.]}         \\
\hline
   $\lambda'_{111}$ \,\, & \, 0.056 \, & 0.048
             & 0.003 & $ \beta \beta 0\nu$ decay \hfill~\cite{HIRSCH95} \\
   $\lambda'_{112}$ \,\, & \, 0.14  \, & 0.12
             & 0.05  & CC-univ. \hfill~\cite{BARGER89} \\
   $\lambda'_{113}$ \,\, & \, 0.18  \,  & 0.15
             & 0.05  & CC-univ. \hfill~\cite{BARGER89} \\
   $\lambda'_{121}$ \,\, & \, 0.058 \, & 0.048
             & 0.6   & $D^+$ decays  \hfill~\cite{BHATTA95}  \\
   $\lambda'_{122}$ \,\, & \, 0.19  \, & 0.16
             & 0.04  & $\nu_e$-mass \hfill~\cite{GODBOL87} \\
   $\lambda'_{123}$ \,\, & \, 0.30  \, & 0.26
             & 0.6   & $D^+$ decays  \hfill~\cite{BHATTA95} \\
   $\lambda'_{131}$ \,\, & \, 0.06  \, & 0.05
             & 0.7   & $\Gamma_h/\Gamma_e\mid^Z$ \hfill~\cite{BHATTB95} \\
   $\lambda'_{132}$ \,\, & \, 0.22  \,  & 0.19
             & 0.7   & $\Gamma_h/\Gamma_e\mid^Z$ \hfill~\cite{BHATTB95} \\
   $\lambda'_{133}$ \,\, & \, 0.55  \,  & 0.48
             & 0.002 & $\nu_e$-mass \hfill~\cite{GODBOL87}\\
\hline
\end{tabular*}
\end{table}
The misidentified DIS CC or photoproduction background in event topologies
of type (C) are suppressed by a combination of loose lepton ($e$ or $\mu$'s)
identification and/or stringent \PTmiss\ requirements
($\PTmiss > 15$ or $25 \GeV$).  
A good signal detection efficiency is obtained in each possible channel,
e.g. ranging from 25\% to 80\% at $M_{\tilde{q}} = 150 \GeV$
depending on the event topologies and model parameters.

Except for a slight ($\sim 2.4 \sigma$) excess around
$M_{\tilde{q}} = 70 \GeV$~\cite{H1RPSUSY,WARSAW96} for ``right'' sign events 
in (B), an excellent agreement is found between observations and background
expectations in all channels and exclusion limits at 95\% CL are derived.
The results for $\lambda'_{1j1}$ combining all contributing channels are 
shown in Fig.~\ref{fig:limplan} ({\it top}) for $M_{\chi_1^0} = 40 \GeV$. 
The existence of squarks with \Rp\ Yukawa couplings $\lambda'_{1j1}$ is
excluded for masses up to
$\sim 220 \GeV$ ($240 \GeV$ for $M_{\chi_1^0} \simeq 160 \GeV$) 
for coupling values 
$\lambda'_{1j1} \gtrsim \sqrt{ 4 \pi \alpha_{em}} $.
The Fig.~\ref{fig:limplan} ({\it bottom}) shows the excluded domains in the 
$M_2$ versus $\mu$ plane for $M_{\tilde{q}} = 150 \GeV$ and $\lambda'$ values
near (or below) the sensitivity limit which is obtained at fixed
$M_{\chi_1^0} = 40 \GeV$.
 
%
%
To our knowledge,
no direct searches in the $\Rp$-SUSY framework with $\lambda' \neq 0$
have been performed yet by $e^+e^-$ and $p\bar{p}$ experiments.
It is likely that in $e^+e^-$ collisions at LEP 200 where squarks are 
dominantly pair produced through gauge couplings, 
squarks masses up to near the
kinematical limit of $\sqrt{s_{e^+e^-}}/2 \simeq 90 \GeV$ can be
reached within few years.
In $p\bar{p}$ collisions at the TEVATRON, squarks can be produced in pairs 
or in association with gluinos.
Given the current mass reach for scalar leptoquark and MSSM searches in 
D0 and CDF experiments~\cite{D0CDF},
it is likely that $M_{\tilde{q}} \gtrsim 200 \GeV$ can be probed through
most of the MSSM parameter space even for $M_{\tilde{g}} > M_{\tilde{q}}$. 
Such a mass reach might be provided already by di-lepton data alone as
was inferred in ref.~\cite{ROY96}. 

%
%
From the analysis of the $\lambda'_{1j1}$ case involving the
$\bar{\tilde{d}}_R$ and  $\tilde{u}_L$ squarks, limits can be
deduced on the $\lambda'_{1jk}$ by folding in the proper parton
densities. Such limits are given in Table~\ref{tab:limtab} at
$M_{\tilde{q}}=150 \GeV$.
%
%
Also quoted in this table are the most severe indirect limits
existing. 
%
\begin{figure}[htb]
\vspace{-1.0cm}
 
\hspace*{-0.7cm} \epsfxsize=8.4cm \epsffile{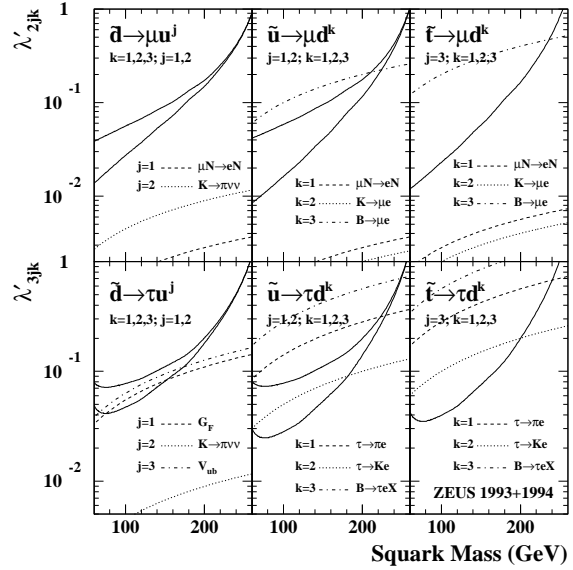}
 
\vspace{-1.0cm}

 \caption[]{ \label{fig:limprod}
   {\small Exclusion upper limits on $\lambda'_{2jk}$ ({\it top})
           and $\lambda'_{3jk}$ ({\it bottom}) at 95\% CL for squarks produced 
           through $\lambda'_{1jk}$ (regions above the curves excluded).
           Assuming a pure $\tilde{\gamma}$ LSP, the $\tilde{d}^k$ limits are 
           derived for $\lambda'_{11k} = \lambda'_{ijk}$ with $i=2,3$ while
           the $\tilde{u}^j$ and $\tilde{t}$ limits are derived for
           $\lambda'_{1j1} = \lambda'_{ijk}$ with $i=2,3$.
           The $\tilde{t}$ limits assumes a mass mixing of
           $\cos \theta_t = 0.5$ and $M_{\tilde{t}} < M_t$.
           The upper (lower) solid curves are the results obtained
           for $M_{\tilde{\gamma}} \ll M_{\tilde{q}}$
           ($M_{\tilde{\gamma}} \sim M_{\tilde{q}}$).
           Indirect limits from other experiments~\cite{DAVIDSON} are
           shown as dashed and dotted curves. }}
\end{figure}
The most stringent concerns $\lambda'_{1jk}$ with
$j=k$ and come either from the non-observation of neutrinoless
double-beta decay ($j=k=1$) or from constraints on the $\nu_e$ mass
($j=k=2,3$).
The limit from $ \beta \beta 0\nu$ decay depends on the gluino
mass~\cite{HIRSCH95} and is given here for $M_{\tilde{g}}= 1 \TeV$.	
The limits on $\lambda'_{11k}$ from CC universality~\cite{BARGER89}
affect only the $\tilde{d}_R^k$.
The limits inferred~\cite{GODBOL87} from $M_{\nu_e}$ constraints
assume slepton and squark mass degeneracy and are valid only~\cite{BHATTA95}
for $\lambda'_{1jk}$ with $j=k$.

%
%
In the presence of two simultaneously non-vanishing Yukawa couplings
(e.g. $\lambda'_{1jk}$ and $\lambda'_{ijk}$ with $i \neq 1$), resonant
$\tilde{q}$ production at HERA can be directly followed by a lepton flavor
violation (LFV) decay leading to $\mu + {\mbox{jet}}$ or $\tau + {\mbox{jet}}$
signatures.
Relevant analyses for these quasi-background free channels
have been performed by the H1~\cite{H1LQ95} and ZEUS~\cite{ZEUSLFV}
collaborations.
Exclusion limits in the context of $\Rp$-SUSY have been derived by
ZEUS~\cite{ZEUSLFV} based on $ L = 3.8 \picob$ of data and
taking into account the branching ratio for gauge decays
involving a pure $\tilde{\gamma}$ LSP (more unfavorable cases of mixed
gaugino-higgsino mass eigenstates have not been considered)
and assuming that only one type of squark is produced.
The results are shown in Fig.~\ref{fig:limprod}.
A sensitivity comparable or better than existing indirect LFV
limits~\cite{DAVIDSON} is obtained at $M_{\tilde{q}} \simeq 150 \GeV$
for some coupling combinations.
Squarks masses up to $\sim 210 \GeV$ (depending on the LFV couplings)
are excluded for $\lambda' \gtrsim \sqrt{ 4 \pi \alpha_{em}} $.
 
\section{CONCLUSION}
\label{sec:concl}

Selectrons and squarks of the Minimal Supersymmetric Standard Model
were searched through pair production at HERA and masses up to $65 \GeV$
were excluded in a new region at large $M_2$ and $\mu \ll 0$. 

Squarks of $R$-parity violating supersymmetry were searched through direct
resonant production via Yukawa couplings $\lambda'_{1jk}$ by the H1 and
ZEUS experiments at HERA.

Assuming that one of the $\lambda'_{1jk}$ dominates, three families of event
topologies were identified for $R$-parity violating and gauge decays of 
squarks.
It was found that squark decays via a $\lambda'$ coupling into $l + q$
final states dominate only at largest accessible masses, while elsewhere
squarks undergo mainly gauge decays into a quark and a 
(generally) unstable gaugino-higgsino.
No significant evidence for the production of squarks was found in any of 
the channels and mass dependent limits on the couplings were derived.
The existence of first generation squarks with masses up to 
$220 \rightarrow 240 \GeV$ (depending on the MSSM parameter values) are
excluded by H1 at $95\%$ confidence level for 
$\lambda'_{1j1} \gtrsim \sqrt{4 \pi \alpha_{em}} $.
These limits extend beyond the current reach at other existing colliders.
At $M_{\tilde{q}} = 150 \GeV$, the upper limits obtained 
for the four couplings $\lambda'_{1jk}$ with $j \neq 1$ and $j \neq k$
are comparable or better than the most stringent indirect limits existing .
 
Considering a combination $\lambda'_{1jk} \times \lambda'_{ijk} \neq 0$ 
with $i \neq 1$,
squarks can undergo direct lepton flavor violation decays leading to
$\mu + {\mbox{jet}} + X$ or $\tau + {\mbox{jet}} + X$ signatures.
For some coupling products, a sensitivity comparable or better than the most
stringent indirect limit existing was obtained by ZEUS in the case of
a ``pure'' $\tilde{\gamma}$ LSP.
 

 
%
\end{document}